\documentclass[12pt,english]{article}
\usepackage[T1]{fontenc}
\usepackage[latin9]{inputenc}
\usepackage{geometry}
\usepackage{amsmath}
\usepackage{amsthm}
\usepackage{amssymb}
\usepackage{graphicx}
\usepackage{setspace}
\usepackage[authoryear]{natbib}
\makeatletter
  \theoremstyle{definition}
  \newtheorem{defn}{\protect\definitionname}
  \theoremstyle{plain}
  \newtheorem{lem}{\protect\lemmaname}
  \theoremstyle{plain}
  \newtheorem{cor}{\protect\corollaryname}
  \theoremstyle{remark}
  \newtheorem{rem}{\protect\remarkname}
\theoremstyle{plain}
\newtheorem{thm}{\protect\theoremname}

\makeatother

\usepackage{babel}
  \providecommand{\definitionname}{Definition}
  \providecommand{\lemmaname}{Lemma}
  \providecommand{\remarkname}{Remark}
\providecommand{\corollaryname}{Corollary}
\providecommand{\theoremname}{Theorem}

\begin{document}

\title{{\Large{}A Canonical Hidden-Variable Space}\thanks{We are grateful to Samson Abramsky, Bob Coecke, Amanda Friedenberg,
Barbara Rifkind, Gus Stuart, and Noson Yanofsky for valuable conversations,
to John Asker, Axelle Ferrière, Andrei Savochkin, participants at
the workshop on Semantics of Information, Dagstuhl, June 2010, participants
at the conference on Advances in Quantum Theory, Linnaeus University,
Växjö, June 2010, and a referee for useful input, and to the NYU Stern
School of Business, NYU Shanghai, and J.P. Valles for financial support.}}

\author{Adam Brandenburger\thanks{Stern School of Business, Tandon School of Engineering, NYU Shanghai,
New York University, New York, NY 10012, U.S.A., adam.brandenburger@stern.nyu.edu,
http://www.adambrandenburger.com}\hspace{7bp}and H. Jerome Keisler\thanks{Department of Mathematics, University of Wisconsin-Madison, Madison,
WI 53706, U.S.A., keisler@math.wisc.edu, http://www.math.wisc.edu/\textasciitilde{}keisler}}

\date{First version: January 15, 2014\\
This version: August 1, 2017}
\maketitle
\begin{abstract}
The hidden-variable question is whether or not various properties
\textemdash{} randomness or correlation, for example \textemdash{}
that are observed in the outcomes of an experiment can be explained
via introduction of extra (hidden) variables which are unobserved
by the experimenter. The question can be asked in both the classical
and quantum domains. In the latter, it is fundamental to the interpretation
of the quantum formalism (Bell, 1964, Kochen and Specker, 1967, and
others). In building a suitable mathematical model of an experiment,
the physical set-up will guide us on how to model the observable variables
\textemdash{} i.e., the measurement and outcome spaces. But, by definition,
we cannot know what structure to put on the hidden-variable space.
Nevertheless, we show that, under a measure-theoretic condition, the
hidden-variable question can be put into a canonical form. The condition
is that the $\sigma$-algebras on the measurement and outcome spaces
are countably generated. An argument using a classical result on isomorphisms
of measure algebras then shows that the hidden-variable space can
always be taken to be the unit interval equipped with the Lebesgue
measure on the Borel sets.
\end{abstract}

\section{Introduction}

\thispagestyle{empty}Consider an experiment in which Alice can make
one of several measurements on her part of a certain system and Bob
can make one of several measurements on his part of the system. Each
pair of measurements (one by Alice and one by Bob) leads to a pair
of outcomes (one for Alice and one for Bob). We keep track of the
frequency distribution of the different pairs of outcomes that arise.
This situation can be abstracted to an \textbf{empirical model}, which,
for each pair of measurements, specifies a probability measure on
pairs of outcomes.

An associated \textbf{hidden-variable model} is obtained by starting
with the empirical model and then appending to it extra variables
that are assumed to be present in a more complete theory of how the
data are generated. The uses of hidden-variable analysis include:
(i) seeing if a deterministic account can be given of the observed
data, and (ii) seeing if a common-cause account can be given of correlations
in the observed data.

Arguably, the most famous context for hidden-variable analysis is
quantum mechanics (QM). Starting with von Neumann (1932), and including,
most famously, Einstein, Podolsky, and Rosen (1935), Bell (1964),
and Kochen and Specker (1967), a vast literature has grown up around
the question of whether or not a hidden-variable formulation of QM
is possible. The watershed no-go theorems of Bell and Kochen-Specker
give conditions under which the answer is no.

Hidden variables are variables above and beyond those which are part
of the actual experiment, and are therefore unobserved. This poses
a question: \textit{What can one assume about the structure of the
space on which a hidden variable lives?} Choosing a good empirical
model includes choosing measurement and outcome spaces that incorporate
appropriate physical features (say, discreteness, connectedness, or
other features). But, since it is unobserved, there is no such guide
to choosing a hidden-variable space.

This may or may not be a serious obstacle. The question often under
study is whether or not a hidden-variable model exists that exhibits
desired properties such as determinism or common-cause correlation.
For a positive answer, we may be satisfied with showing that, at least
for a certain choice of hidden-variable space, such a model exists.
But, some of the most important results \textemdash{} including the
Bell and Kochen-Specker theorems \textemdash{} are negative answers,
asserting that no hidden-variable model with certain properties exists.
For a non-existence result to be definitive, we need to search over
all (not just some) hidden-variable spaces.

Our main result is that there is a \textbf{canonical hidden-variable
model}. To be more specific, fix an empirical model. Suppose there
is an associated hidden-variable model that yields, for each pair
of measurements, the same probability measure over joint outcomes.
We will say that the hidden-variable model \textbf{realizes} the empirical
model. We want to know if we can put this hidden-variable model into
a canonical form. More than this, to be useful, such a canonical hidden-variable
model must preserve properties \textemdash{} determinism, common-cause
correlation, etc. \textemdash{} satisfied by the original hidden-variable
model.

We show that, under a measure-theoretic condition, such a canonical
model exists: \emph{If the $\sigma$-algebras on the measurement and
outcome spaces are countably generated, then the hidden-variable space
can always be taken to be the unit interval equipped with the Lebesgue
measure on the Borel sets}. Note that if a probability space has a
countably generated $\sigma$-algebra, then the associated probability
algebra is separable. The key to our result is the classical theorem
that any two separable atomless probability algebras are isomorphic.
This theorem can be found in Carath\'eodory (1939) and Halmos and
von Neumann (1942). It is also a special case of Maharam's Theorem
(Maharam, 1942).

\section{Empirical and Hidden-Variable Models}

Alice has a space of possible measurements, which is a measurable
space $(Y_{a},\mathcal{Y}_{a})$, and a space of possible outcomes,
which is a measurable space $(X_{a},\mathcal{X}_{a})$. Likewise,
Bob has a space of possible measurements, which is a measurable space
$(Y_{b},\mathcal{Y}_{b})$, and a space of possible outcomes, which
is a measurable space $(X_{b},\mathcal{X}_{b})$. We will restrict
attention to bipartite systems. (We comment later on the extension
to more than two parts.) There is also a hidden-variable space, which
is an unspecified measurable space $(\Lambda,\mathcal{L})$. Write
\begin{align*}
(X,\mathcal{X}) & =(X_{a},\mathcal{X}_{a})\otimes(X_{b},\mathcal{X}_{b}),\\
(Y,\mathcal{Y}) & =(Y_{a},\mathcal{Y}_{a})\otimes(Y_{b},\mathcal{Y}_{b}),\\
\Psi & =(X,\mathcal{X})\otimes(Y,\mathcal{Y}),\\
\Omega & =(X,\mathcal{X})\otimes(Y,\mathcal{Y})\otimes(\Lambda,\mathcal{L}).
\end{align*}

\begin{defn}
An \textbf{empirical model} is a probability measure $e$ on $\Psi$.
\end{defn}
We see that an empirical model describes an experiment in which the
pair of measurements $y=(y_{a},y_{b})\in Y$ is randomly chosen according
to the probability measure ${\rm marg}_{Y}e$, and $y$ and the joint
outcome $x=(x_{a},x_{b})\in X$ are distributed according to $e$.
\begin{defn}
A \textbf{hidden-variable model} is a probability measure $p$ on
$\Omega$.
\end{defn}
\,
\begin{defn}
We say that a hidden-variable model $p$ \textbf{realizes} an empirical
model $e$ if $e={\rm marg}_{\Psi}p$. We say that two hidden-variable
models are (\textbf{realization-})\textbf{equivalent} if they realize
the same empirical model.
\end{defn}

\section{Preliminaries}

Throughout, we use the following two conventions. First, when $p$
is a probability measure on a product space $(X,\mathcal{X})\otimes(Y,\mathcal{Y})$
and $q={\rm marg}_{X}p$, then for each $J\in\mathcal{X}$ we write

\[
p(J)=p(J\times Y)=q(J),
\]
and for each $q$-integrable $f:X\rightarrow\mathbb{R}$ we write

\[
\int_{J}f(x)\,dp=\int_{J\times Y}f(x)\,dp=\int_{J}f(x)\,dq.
\]
Thus, in particular, a statement holds for $p$-almost all $x\in X$
if and only if it holds for $q$-almost all $x\in X$.

Second, when $p$ is a probability measure on a product space $(X,\mathcal{X})\otimes(Y,\mathcal{Y})\otimes(Z,\mathcal{Z})$,
and $J\in\mathcal{X}$, we let $p[J||\mathcal{Z}]$ be the function
from $Z$ into $[0,1]$ such that
\[
p[J||\mathcal{Z}]_{z}=p[J\times Y\times Z|\{X\times Y,\emptyset\}\otimes\mathcal{Z}]_{(x,y,z)}={\rm E}[1_{J\times Y\times Z}|\{X\times Y,\emptyset\}\otimes\mathcal{Z}].
\]
Note that $\{X\times Y,\emptyset\}$ is the trivial $\sigma$-algebra
over $X\times Y$.

We use similar notation for (finite) products with factors to the
left of $(X,\mathcal{X})$ or to the right of $(Z,\mathcal{Z})$.
Note that if $q={\rm marg}_{X\times Z}p$, then $q[J||\mathcal{Z}]=p[J||\mathcal{Z}]$.
We also use the analogous notation for expected values of random variables:
Given an integrable function $f:X\rightarrow\mathbb{R}$, we write
${\rm E}[f||\mathcal{Z}]$ for the conditional expectation ${\rm E}[f\circ\pi|\{X\times Y,\emptyset\}\otimes\mathcal{Z}]$where
$\pi$ is the projection from $X\times Y\times Z$ to $X$.
\begin{lem}
\label{l-cond} The mapping $z\mapsto p[J||\mathcal{Z}]_{z}$ is the
$p$-almost surely unique $\mathcal{Z}$-measurable function $f:Z\rightarrow[0,1]$
such that for each set $L\in\mathcal{Z}$,
\[
\int_{L}f(z)\,dp=p(J\times L).
\]
\end{lem}
\begin{proof}
Existence: Let $f(z)=p[J||\mathcal{Z}]_{z}$. Using the definition
of $p[J||\mathcal{Z}]$, we see that
\begin{multline*}
\int_{L}f(z)\,dp=\int_{X\times Y\times L}{\rm E}[1_{J\times Y\times Z}|\{X\times Y,\emptyset\}\otimes\mathcal{Z}]\,dp=\\
\int_{X\times Y\times L}1_{J\times Y\times Z}\,dp=p((X\times Y\times L)\cap(J\times Y\times Z))=p(J\times L),
\end{multline*}

Uniqueness: Let $f$ and $g$ be two such functions and let $L=\{z:f(z)<g(z)\}$.
Then $L\in\mathcal{Z}$. If $p(J)=0$, then $f(z)=g(z)=0$, $p$-almost
surely. Next suppose $p(J)>0$. If $p(J\times L)>0$, then $p(L)>0$,
and
\[
0<\int_{L}g(z)\,dp-\int_{L}f(z)\,dp=\int_{L}(g(z)-f(z))\,dp=0,
\]
a contradiction. Therefore $p(J\times L)=0$, so $p(L)=0$ and hence
$f(z)\ge g(z)$, $p$-almost surely. Similarly, $g(z)\ge f(z)$, $p$-almost
surely, so $f(z)=g(z)$, $p$-almost surely.
\end{proof}
\begin{cor}
\label{c-marg} Let $q$ be the marginal of $p$ on $X\times Z$.
Then, for each $J\in\mathcal{X}$, we have $p[J||\mathcal{Z}]=q[J||\mathcal{Z}]$,
$q$-almost surely.
\end{cor}
Given probability measures $p$ on $(X,\mathcal{X})\otimes(Y,\mathcal{Y})$
and $r$ on $(Y,\mathcal{Y})$, we say that $p$ is an \textbf{extension}
of $r$ if $r={\rm marg}_{Y}p$. We say that two probability measures
$p$ and $q$ on $(X,\mathcal{X})\otimes(Y,\mathcal{Y})$ \textbf{agree
on} $Y$ if ${\rm marg}_{Y}p={\rm marg}_{Y}q$.

\section{Properties of Hidden-Variable Models}

Whenever we write an equation involving conditional probabilities,
it will be understood to mean that the equation holds $p$-almost
surely. All expressions below which are given for Alice have counterparts
for Bob, with $a$ and $b$ interchanged.

The properties of hidden-variable models we now list are all stated
for infinite measurement and outcome spaces. In the sources we give,
these properties were stated for the finite case only.

We begin with Bell (1964) locality:
\begin{defn}
The hidden-variable model $p$ satisfies \textbf{locality} if for
every $J_{a}\in\mathcal{X}_{a}$, $J_{b}\in\mathcal{X}_{b}$, we have
\[
p[J_{a}\times J_{b}||\mathcal{Y}\otimes\mathcal{L}]=p[J_{a}||\mathcal{Y}_{a}\otimes\mathcal{L}]\times p[J_{b}||\mathcal{Y}_{b}\otimes\mathcal{L}].
\]
\end{defn}
The next two properties come from Jarrett (1984) (and were given these
names by Shimony, 1986).
\begin{defn}
The hidden-variable model $p$ satisfies \textbf{parameter independence}
if for every $J_{a}\in\mathcal{X}_{a}$ we have
\[
p[J_{a}||\mathcal{Y}\otimes\mathcal{L}]=p[J_{a}||\mathcal{Y}_{a}\otimes\mathcal{L}].
\]
\end{defn}
\,
\begin{defn}
The hidden-variable model $p$ satisfies \textbf{outcome independence}
if for every $J_{a}\in\mathcal{X}_{a}$, $J_{b}\in\mathcal{X}_{b}$,
we have
\[
p[J_{a}\times J_{b}||\mathcal{Y}\otimes\mathcal{L}]=p[J_{a}||\mathcal{Y}\otimes\mathcal{L}]\times p[J_{b}||\mathcal{Y}\otimes\mathcal{L}].
\]
\end{defn}
The name of the next property is due to Dickson (2005).
\begin{defn}
The hidden-variable model $p$ satisfies $\lambda$\textbf{-independence}
if for every event $L\in\mathcal{L}$,
\[
p[L||\mathcal{Y}]_{y}=p(L).
\]
\end{defn}
\,
\begin{rem}
\label{r-product} We observe:
\end{rem}
\begin{enumerate}
\item The $\lambda$-independence property for $p$ depends only on ${\rm marg}_{Y\times\Lambda}p$.
\item Any hidden-variable model $p$ such that $\Lambda$ is a singleton
satisfies $\lambda$-independence.
\end{enumerate}
\,

By Remark \ref{r-product}, we have:
\begin{lem}
\label{l-indep} The following are equivalent:
\end{lem}
\begin{enumerate}
\item $p$ satisfies $\lambda$-independence.
\item ${\rm marg}_{Y\times\Lambda}p={\rm marg}_{Y}p\otimes{\rm marg}_{\Lambda}p$.
\item The $\sigma$-algebras $\mathcal{Y}$ and $\mathcal{L}$ are independent
with respect to $p$, i.e.,
\[
p(K\times L)=p(K)\times p(L)
\]
for every $K\in\mathcal{Y},L\in\mathcal{L}$.
\end{enumerate}
The distinction between strong and weak determinism in the next two
definitions is from Brandenburger and Yanofsky (2008).
\begin{defn}
The hidden-variable model $p$ satisfies \textbf{strong determinism}
if for each $J_{a}\in\mathcal{X}_{a}$, we have
\[
p[J_{a}||\mathcal{Y}_{a}\otimes\mathcal{L}]_{(y_{a},\lambda)}\in\{0,1\}.
\]
\end{defn}
\,
\begin{defn}
The hidden-variable model $p$ satisfies \textbf{weak determinism}
if for every $J_{a}\in\mathcal{X}_{a}$, $J_{b}\in\mathcal{X}_{b}$,
we have
\[
p[J_{a}\times J_{b}||\mathcal{Y}\otimes\mathcal{L}]_{(y,\lambda)}\in\{0,1\}.
\]
\end{defn}
\hspace{40bp}\includegraphics[scale=0.5]{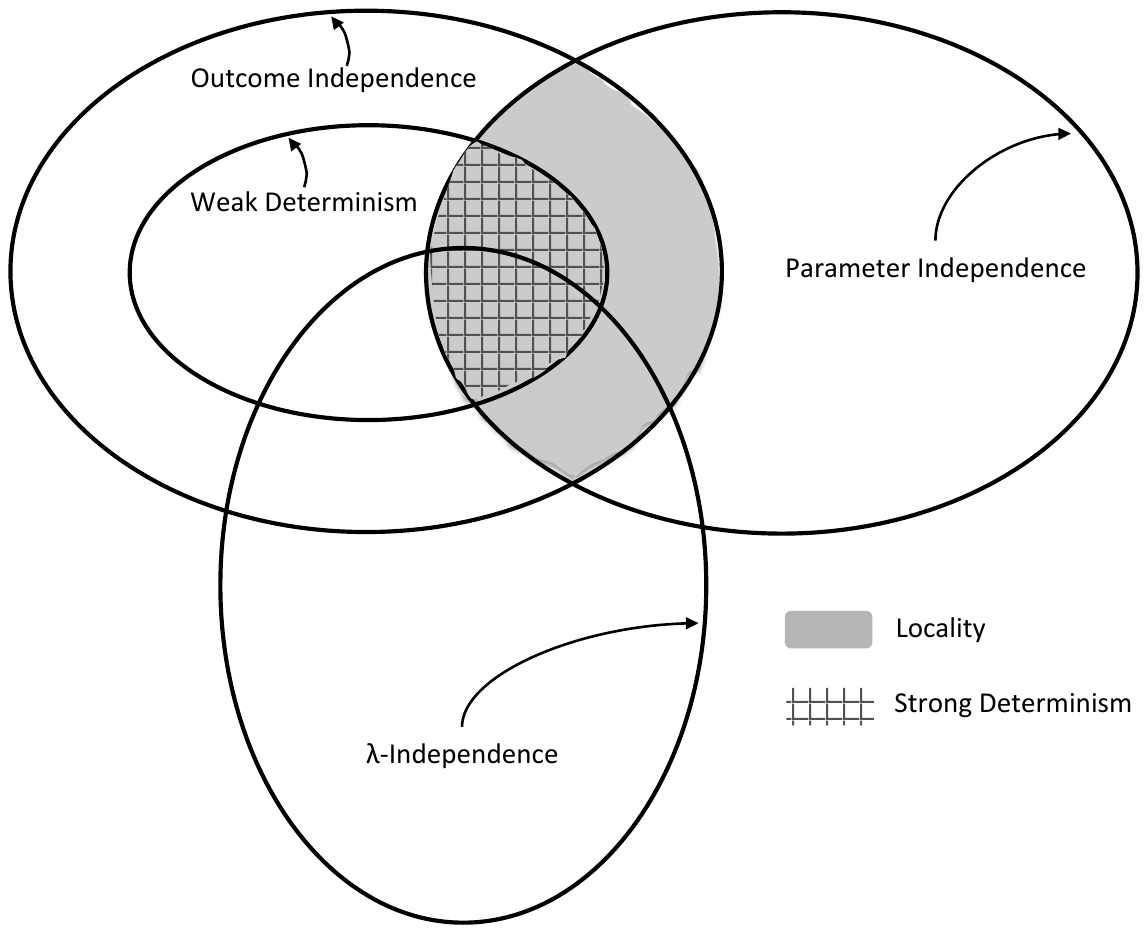}

\hspace{200bp}Figure 1

\bigskip{}

In Brandenburger and Keisler (2016), we establish how the various
properties we have listed above are related, in the case that the
outcome spaces $X_{b}$ and $X_{b}$ are finite. The Venn diagram
above summarizes these relationships. The same relationships hold
for general (infinite) $X_{a}$ and $X_{b}$.

To see this, we first show that the relationships automatically carry
over to the case that the $\sigma$-algebras $\mathcal{X}_{a}$ and
$\mathcal{X}_{b}$ are finite. In this case, let $X_{a}^{\prime}$
be the finite set of atoms of $\mathcal{X}_{a}$ and let $\mathcal{X}_{a}^{\prime}$
be the power set of $X_{a}^{\prime}$. Define $X_{b}^{\prime}$ and
$\mathcal{X}_{b}^{\prime}$ similarly. Then the space
\[
\Omega^{\prime}=(X^{\prime},\mathcal{X}^{\prime})\otimes(Y,\mathcal{Y})\otimes(\Lambda,\mathcal{L})
\]
behaves exactly like the space $\Omega$. For each hidden-variable
model $p$, let $p^{\prime}$ be the hidden-variable model on $\Omega^{\prime}$
such that $p^{\prime}(\{x^{\prime}\}\times K\times L)=p(x^{\prime}\times K\times L)$,
and, for each empirical model $e$, let $e^{\prime}$ be the empirical
model such that $e^{\prime}(\{x^{\prime}\}\times K)=e(x^{\prime}\times K)$.
It is clear that $p$ realizes $e$ if and only if $p^{\prime}$ realizes
$e^{\prime}$, that $p^{\prime}$ has the same hidden-variable space
as $p$, and that each of the properties we have listed holds for
$p^{\prime}$ if and only if it holds for $p$. It follows that the
same Venn diagram of relationships holds for the case that the $\sigma$-algebras
$\mathcal{X}_{a}$ and $\mathcal{X}_{b}$ are finite.

Next, we extend this observation to the case of general $\mathcal{X}_{a}$
and $\mathcal{X}_{b}$.
\begin{defn}
Let $\mathcal{F}_{a}$ and $\mathcal{F}_{b}$ be finite subalgebras
of $\mathcal{X}_{a}$ and $\mathcal{X}_{b}$, respectively, and let
$\mathcal{F}=\mathcal{F}_{a}\otimes\mathcal{F}_{b}$. For each empirical
model $e$, the\textbf{ finite restriction} $e\upharpoonright\mathcal{F}$
of $e$ to $\mathcal{F}$ is the restriction of $e$ to the $\sigma$-algebra
$\mathcal{F}\otimes\mathcal{Y}$. For each hidden-variable model $p$,
the \textbf{finite restriction} $p\upharpoonright\mathcal{F}$ of
$p$ to $\mathcal{F}$ is the restriction of $p$ to the $\sigma$-algebra
$\mathcal{F}\otimes\mathcal{Y}\otimes\mathcal{L}$.
\end{defn}
The following lemma is an easy consequence of the definitions.
\begin{lem}
We have:
\end{lem}
\begin{enumerate}
\item $p$ realizes $e$ if and only if every finite restriction of $p$
realizes the corresponding finite restriction of $e$.
\item $p$ satisfies locality, parameter independence, outcome independence,
$\lambda$-independence, strong determinism, or weak determinism if
and only if every finite restriction of $p$ satisfies the corresponding
property.
\end{enumerate}
\,

From this it follows that the same Venn diagram of relationships holds
for the case of general $\mathcal{X}_{a}$ and $\mathcal{X}_{b}$.

Finally in this section, we note that the definitions and relationships
we have given extend immediately to systems with more than two parts,
except that parameter independence must now be stated in terms of
sets of parts instead of individual parts.

\section{A Canonical Hidden-Variable Space}

Given a hidden-variable model $p$, we let $\ell={\rm {marg}}_{\Lambda}p$
and call the probability space $(\Lambda,\mathcal{L},\ell)$ the \textbf{hidden-variable
space of} $p$. We now state and prove our result on the existence
of a canonical hidden-variable space.
\begin{thm}
\label{t-canonical} Assume that the $\sigma$-algebras $\mathcal{X}$
and $\mathcal{Y}$ are countably generated. Then every hidden-variable
model $p$ is realization-equivalent to a hidden-variable model $\overline{p}$
with hidden-variable space $([0,1],\mathcal{U},u)$, where $\mathcal{U}$
is the set of Borel subsets of $[0,1]$ and $u$ is Lebesgue measure
on $\mathcal{U}$. Moreover, for each of the properties of parameter
independence and outcome independence (and, therefore, locality),
$\lambda$-independence, strong determinism, and weak determinism,
if $p$ has the property then so does $\overline{p}$.
\end{thm}
To begin the proof of Theorem 1, we may assume without loss of generality
that $p$ has an atomless hidden-variable model, because the product
$p\otimes u$ of $p$ with the Lebesgue unit interval is realization-equivalent
to $p$ and has the atomless hidden-variable model $(\Lambda,\mathcal{L},\ell)\otimes([0,1],\mathcal{U},u)$,
and if $p$ has any of the five properties above, then so does $p\otimes u$.

Since $\mathcal{X}$ and $\mathcal{Y}$ are countably generated, $\mathcal{X}_{a},\mathcal{X}_{b},\mathcal{Y}_{a}$,
and $\mathcal{Y}_{b}$ are countably generated. Let $\mathcal{X}_{a}^{0}$
be a countable subset of $\mathcal{X}_{a}$ that generates the $\sigma$-algebra
$\mathcal{X}_{a}$, and let $\mathcal{Y}_{a}^{0}$ be a countable
subset of $\mathcal{Y}_{a}$ that generates the $\sigma$-algebra
$\mathcal{Y}_{a}$. Define $\mathcal{X}_{b}^{0}$ and $\mathcal{Y}_{b}^{0}$
similarly. Let $\mathcal{U}_{0}$ be the family of all open subintervals
of $[0,1]$ with rational endpoints. Let 
\[
(A_{an},A_{bn},B_{an},B_{bn},C_{n})
\]
be an enumeration of the countable set $\mathcal{X}_{a}^{0}\times\mathcal{X}_{b}^{0}\times\mathcal{Y}_{a}^{0}\times\mathcal{Y}_{b}^{0}\times\mathcal{U}_{0}$.
The Cartesian products $A_{an}\times A_{bn}\times B_{an}\times B_{bn}\times C_{n}$
generate the $\sigma$-algebra $\mathcal{X}\otimes\mathcal{Y}\otimes\mathcal{U}$.
For each $n$, let $f_n\colon\Lambda\to[0,1]$ be a representative
of the family of $p$-almost surely unique functions
\[
\lambda\mapsto p[A_{an}\times A_{bn}\times B_{an}\times B_{bn}||\mathcal{L}]_{\lambda},
\]
and let $D_n=f_n^{-1}(C_n)$.

To continue the proof, we need the following lemma.
\begin{lem}
There is a countably generated $\sigma$-algebra $\mathcal{D}\subseteq\mathcal{L}$
such that each of the sets $D_{n}$ belongs to $\mathcal{D}$, and
the restriction of $\ell$ to $\mathcal{D}$ is atomless.
\end{lem}
\begin{proof}
Since $\ell$ is atomless, it follows from a result of Sierpinski
(1922) that for each set $L\in\mathcal{L}$ there is a set $L^{\prime}\in\mathcal{L}$
such that $L^{\prime}\subseteq L$ and $\ell(L^{\prime})=\ell(L)/2$.
Then, by the Axiom of Choice, there is a function $F:\mathcal{L}\rightarrow\mathcal{L}$
such that for each $L\in\mathcal{L}$, $F(L)\subseteq L$ and $\ell(F(L))=\ell(L)/2$.
Let $\mathcal{E}_{0}$ be the algebra of subsets of $\Lambda$ generated
by $\{D_{n}:n\in\mathbb{N}\}$. For each $m\in\mathbb{N}$, let $\mathcal{E}_{m+1}$
be the algebra of subsets of $\Lambda$ generated by $\mathcal{E}_{m}\cup\{F(L):L\in\mathcal{E}_{m}\}$.
Let $\mathcal{E}=\bigcup_{m}\mathcal{E}_{m}$ and let $\mathcal{D}$
be the $\sigma$-algebra generated by $\mathcal{E}$. Clearly, each
$D_{n}$ belongs to $\mathcal{D}$, and $\mathcal{E}$ is countable,
so $\mathcal{D}$ is countably generated.

We show that the restriction of $\ell$ to $\mathcal{D}$ is atomless.
Let $\mathcal{D}^{\prime}$ be the set of all $D\in\mathcal{D}$ that
can be approximated by sets in $\mathcal{E}$ with respect to $\ell$,
that is,
\[
\mathcal{D}^{\prime}=\{D\in\mathcal{D}:(\forall r>0)(\exists E\in\mathcal{E})\ell(E\triangle D)<r\}.
\]

It is clear that $\mathcal{E}\subseteq\mathcal{D}^{\prime}$, and
that $\mathcal{D}^{\prime}$ is closed under finite unions and intersections.
The set $\mathcal{D}^{\prime}$ is also closed under unions of countable
chains, because if $L_{n}\in\mathcal{D}^{\prime}$ and $L_{n}\subseteq L_{n+1}$
for each $n$, and $L=\bigcup_{n}L_{n}$, then for each $r>0$ there
exists $n\in\mathbb{N}$ and $E\in\mathcal{E}$ such that $\ell(L\triangle L_{n})<r/2$
and $\ell(E\triangle L_{n})<r/2$. Therefore $\ell(L\triangle E)<r$,
so $L\in\mathcal{D}^{\prime}$. It follows that $\mathcal{D}^{\prime}=\mathcal{D}$.
Now suppose $D\in\mathcal{D}$, $\ell(D)>0$, and $r>0$. Then $D\in\mathcal{D}^{\prime}$,
so there is a set $G\in\mathcal{E}$ such that $\ell(D\triangle G)<r$.
We have $F(G)\in\mathcal{E}$, $\ell(F(G))=\ell(G)/2$, and $F(G)\subseteq G$.
Then $D\cap F(G)\in\mathcal{D}$, and, by taking $r$ small enough,
we can guarantee that $\ell(D)>\ell(D\cap F(G))>0$. This shows that
the restriction of $\ell$ to $\mathcal{D}$ is atomless, and proves
the lemma.
\end{proof}
Continuing the proof of Theorem 1, we know by Carath\'eodory (1939)
and Halmos and von Neumann (1942) (and a special case in Maharam,
1942) that any two separable atomless probability algebras are isomorphic.
Therefore, the measure algebras of $(\Lambda,\mathcal{D},\ell)$ and
$([0,1],\mathcal{U},u)$ are isomorphic. This isomorphism maps the
equivalence class (modulo null sets) of each $D\in\mathcal{D}$ to
the equivalence class of a set $h(D)\in\mathcal{U}$ such that $u(h(D))=\ell(D)$.

Let $\overline{p}$ be the probability measure on $\mathcal{X}\otimes\mathcal{Y}\otimes\mathcal{U}$
such that for each $R\in\mathcal{X}$, $S\in\mathcal{Y}$, and $D\in\mathcal{D}$
we have $\overline{p}(R\times S\times h(D))=p(R\times S\times D)$.
It follows that the hidden-variable space of $\overline{p}$ is $([0,1],\mathcal{U},\ell)$.
Moreover, $\overline{p}$ and $p$ agree on $X\times Y$, so $\overline{p}$
is realization-equivalent to $p$. For each $S\in\mathcal{Y}$ and
$D\in\mathcal{D}$ we have
\[
\overline{p}(S)=p(S)\text{, }\thickspace\overline{p}(h(D))=p(D)\text{, }\thickspace\overline{p}(S\times h(D))=p(S\times D).
\]
 Therefore, if $p$ has $\lambda$-independence, then $\overline{p}$
has $\lambda$-independence by Lemma \ref{l-indep}.

The $\sigma$-algebra $\mathcal{D}$ is large enough so that for each
$K\in\mathcal{Y}$, the function $p[J_{a}\times J_{b}\times K||\mathcal{L}]$
is $\mathcal{D}$-measurable. It follows that
\[
p[J_{a}\times J_{b}||\mathcal{Y}\otimes\mathcal{L}]=p[J_{a}\times J_{b}||\mathcal{Y}\otimes\mathcal{D}],
\]

\[
p[J_{a}||\mathcal{Y}_{a}\otimes\mathcal{L}]=p[J_{a}||\mathcal{Y}_{a}\otimes\mathcal{D}],
\]
and
\[
p[J_{b}||\mathcal{Y}_{b}\otimes\mathcal{L}]=p[J_{b}||\mathcal{Y}_{b}\otimes\mathcal{D}].
\]

From the definition of $\overline{p}$, one can see that joint distributions
of the functions
\[
p[J_{a}\times J_{b}||\mathcal{Y}\otimes\mathcal{D}]\text{, }\thickspace p[J_{a}||\mathcal{Y}_{a}\otimes\mathcal{D}]\text{, }\thickspace p[J_{b}||\mathcal{Y}_{b}\otimes\mathcal{D}]
\]
and
\[
\overline{p}[J_{a}\times J_{b}||\mathcal{Y}\otimes\mathcal{U}]\text{, }\thickspace\overline{p}[J_{a}||\mathcal{Y}_{a}\otimes\mathcal{U}]\text{, }\thickspace\overline{p}[J_{b}||\mathcal{Y}_{b}\otimes\mathcal{U}]
\]
are the same. It follows that for each of the properties of parameter
independence, outcome independence, strong determinism, and weak determinism,
if $p$ has the property then so does $\overline{p}$.

\section*{References}
\begin{enumerate}
\item Abramsky, S., and A. Brandenburger, ``A Unified Sheaf-Theoretic Treatment
of Non-Locality and Contextuality,'' \textit{New Journal of Physics},
13, 2011, 11303.
\item Bell, J., ``On the Einstein-Podolsky-Rosen Paradox,'' \textit{Physics},
1, 1964, 195-200.
\item Brandenburger, A., and H.J. Keisler, ``Fiber Products of Measures
and Quantum Foundations,'' in Chubb J., A. Eskandarian, and V. Harizanov
(eds.), \textit{Logic \& Algebraic Structures in Quantum Computing
\& Information}, Lecture Notes in Logic, Association for Symbolic
Logic/Cambridge University Press, 2016.
\item Brandenburger, A., and N. Yanofsky, ``A Classification of Hidden-Variable
Properties,'' \textit{Journal of Physics A: Mathematical and Theoretical},
41, 2008, 425302.
\item Carath\'eodory, C., ``Die homomorphieen von Somen und die Multiplikation
von Inhaltsfunktionen,'' \textit{Annali della R. Scuola Normale Superiore
di Pisa}, Series 2, 8, 1939, 105-130.
\item Dickson, W.M., \textit{Quantum Chance and Non-Locality: Probability
and Non-Locality in the Interpretations of Quantum Mechanics}, Cambridge
University Press, 2005.
\item Einstein, A., B. Podolsky, and N. Rosen, ``Can Quantum-Mechanical
Description of Physical Reality be Considered Complete?'' \textit{Physical
Review}, 47, 1935, 777-780.
\item Halmos, P., and J. von Neumann, ``Operator Methods in Classical Mechanics,
II,'' \textit{Annals of Mathematics}, 43, 1942, 332-350.
\item Jarrett, J., ``On the Physical Significance of the Locality Conditions
in the Bell Arguments,'' \textit{No\^us}, 18, 1984, 569-589.
\item Kochen, S., and E. Specker, ``The Problem of Hidden Variables in Quantum
Mechanics,'' \textit{Journal of Mathematics and Mechanics}, 17, 1967,
59-87.
\item Maharam, D., ``On Homogeneous Measure Algebras,'' \textit{Proceedings
of the National Academy of Sciences}, 28, 1942, 108-111.
\item Shimony, A., ``Events and Processes in the Quantum World,'' in Penrose,
R., and C. Isham (eds.), \textit{Quantum Concepts in Space and Time},
Oxford University Press, 1986, 182-203.
\item Sierpinski, W., ``Sur les Fonctions d'Ensemble Additives et Continues,''
\textit{Fundamenta Mathematicae}, 3, 1922, 240-246.
\item Von Neumann, J., \textit{Mathematische Grundlagen der Quantenmechanik},
Springer-Verlag, 1932. (Translated as \textit{Mathematical Foundations
of Quantum Mechanics}, Princeton University Press, 1955.)
\end{enumerate}

\end{document}